\documentclass[aps,prl,twocolumn,epsf,epsfig,amsmath,colorlinks=true, superscriptaddress, scitecolor=blue, linkcolor=blue, urlcolor=blue]{revtex4-2}
\usepackage{latexsym}
\usepackage{times}
\usepackage{graphicx}
\usepackage{amsmath}
\usepackage{multirow}
\usepackage{amsthm}
\usepackage{dcolumn}
\usepackage{bm}
\usepackage{ textcomp }
\usepackage{color}
\usepackage{amssymb}
\usepackage{xcolor}
\usepackage{easyReview}
\usepackage{mathdots}
\usepackage{float}
\usepackage{lineno}
\usepackage{todonotes}
\usepackage{array}
\usepackage{orcidlink}
\usepackage{booktabs}
\usepackage{stackengine}
\newcommand{\beq}{\begin{eqnarray}}
\newcommand{\eeq}{\end{eqnarray}}

\begin{document}

\title{2D Helical Twist Controls  Tricritical Point in an Interacting Majorana Chain}

\author{Hekai Zhao$^{1}$ and Philip W. Phillips}

\affiliation{Department of Physics and Anthony J. Leggett Institute of Condensed Matter Theory, University of Illinois at Urbana-Champaign, Urbana, IL 61801, USA}

\date{\today}

\begin{abstract}
We analyze a series of interacting Majorana Fermion chains with finite range pair interactions with coupling strength $g$ that all exhibit a tri-critical point that separates an Ising critical phase from a supersymmetric gapped phase.  We first notice that the interacting models exhibit an even-odd asymmetry depending on the number of sites, $\delta$, over which the interaction ranges.  The even case exhibits competing order, thereby making it numerically untractable while the odd case exhibits an exactly solvable point at $g=-0.5$ where the entanglement entropy vanishes.  By introducing a swirling geometrical twist, we map our 1D $\delta$-range chains to a series of 2D $\delta/2$-width models. Our new 2D models possess a  unique helical boundary condition, constructed from 1D chains with the end of one connected to the start of another.  We propose that the phase transition in the 1D system can be understood as a finite-system size transition in 2D. That is, the $g_c-\delta$ behavior is controlled by a 2D tri-critical universality class at $\delta\to\infty$ limit and is predicted by finite-size scaling theory. 
\end{abstract}

\maketitle

\paragraph{Introduction}

Self conjugacy of Majorana excitations greatly restricts how they interact. For example, in lattice models, there is no Hubbard analogue with Majorana Fermions.  The simplest lattice model with interacting Majoranas places them on four neighboring sites.  In 1D, Rahmani, Zhu, Franz and Affleck (RZFA)\cite{aff1} considered such a model and showed that it has an emergent massive supersymmetric phase.  In particular, the model has a tri-critical Ising (TCI) point which separates a critical phase in the Ising
universality class from a supersymmetric massive phase.  The central charge is $c=7/10$  as opposed to the standard value of  $c=1/2$ of the Ising model.  Further, this tri-criticality appears only when the coupling constant for the interactions is roughly 250 times the kinetic energy of the non-interacting problem.  

A central question that has arisen from this work is why should the coupling constant be so large to generate such universal physics?  Just two years after RZFA, O'brien and Fendley (OF)\cite{obrien1} considered a related model with a slight change. They skipped a site in the interaction term, giving rise to an interaction spread over 5 sites.  Remarkably, this small change shrunk the critical value of the coupling constant down to something on the order of unity, roughly $0.856$\cite{obrien1}. As density-matrix renormalization group (DMRG) is the dominant tool for such 1D systems\cite{aff1,obrien1},  a critical coupling of $O(1)$ greatly reduces the bond dimension required.  Nonetheless, the precise origin of the shift in the coupling constant remains unresolved and is precisely the problem we investigate here.

Part of the attraction of Majorana models is their relationship to spin models via Jordan-Wigner(JW) transformations.  While early field-theoretic arguments based on a $\phi^6$ theory suggested that the TCI phase requires precise multi-parameter fine-tuning \cite{shenkar1}—a constraint evident in many preceding lattice models—Majorana systems provide a more robust platform. In these systems, the self-duality bypasses the need for delicate adjustments, effectively relaxing the conditions required to access the TCI critical point \cite{tci1}.  This is amply illustrated in the work of RZFA\cite{aff1} as they demonstrated a mapping to a 4-spin interacting model and  successfully fixed the transition line exactly through large-scale DMRG calculations. Their analysis of the universal spectrum ratio allowed them to elucidate the `Ising CFT to TCI-CFT to Gapped` phase diagram, confirming the model as a possible realization of the TCI-CFT. 

In this letter, we focus on the odd-integer $\delta$ sector, a choice motivated by the spin structure of the O'Brien-Fendley (OF) model~\cite{obrien1}.
While our DMRG simulations for small $\delta$ reveal a monotonic decrease in the critical coupling $|g_c/t_c|$, the computational cost diverges with interaction range and prevents direct numerical extrapolation.
To overcome this bottleneck, we develop a theoretical framework mapping the 1D variable-range problem to a family of 2D lattice models with width $W \sim \delta$ and helical boundary conditions (HBC).
This mapping allows us to identify the observed $\delta$-dependent shifts as finite-size effects of a single underlying (2+1)D transition.
By analyzing the universality of these effective 2D models, we can asymptotically extrapolate the phase diagram to the infinite-range limit, explaining the robustness of the phase diagram and the drift of the critical point within a common picture.

\paragraph{The Model and Symmetries}

We study a generalized interacting Majorana fermion chain of length $2L$, described by Majorana operators $\gamma_j$ satisfying the Clifford algebra $\{\gamma_i, \gamma_j\} = 2\delta_{ij}$. The Hamiltonian is given by
\begin{equation}
\label{eq:hamiltonian_maj}
H = i t \sum_{j=1}^{2L-1} \gamma_j \gamma_{j+1} + g \sum_{j=1}^{2L-\delta-1} \gamma_j \gamma_{j+1} \gamma_{j+\delta} \gamma_{j+\delta+1},
\end{equation}
where $t$ represents the nearest-neighbor hopping amplitude (which we set to $t=1$ as our energy unit), and $g$ parameterizes the strength of a four-fermion interaction with range $\delta \ge 2$. The standard non-interacting critical free Majorana point corresponds to $g=0$.

To elucidate the connection of $H$ to familiar spin models, we employ the JW transformation. We map the Majorana operators to spin-1/2 degrees of freedom via
\begin{equation}
\gamma_{2j-1} = \left(\prod_{k<j} \sigma_k^x\right) \sigma_j^z, \quad \gamma_{2j} = -i \sigma_j^x \gamma_{2j-1}.
\end{equation}
This mapping enables us to rewrite the nearest-neighbor bilinear terms as the components in the transverse-field Ising terms.  That is, $\gamma_{2j-1}\gamma_{2j} = i\sigma_j^x$ and $\gamma_{2j}\gamma_{2j+1} = i\sigma_j^z \sigma_{j+1}^z$.
Consequently, the free Majorana chain becomes a critical transverse-field Ising model (TFIM):
\begin{equation}
\mathcal{H}_0 = i\sum_j (\gamma_{2j-1}\gamma_{2j} + \gamma_{2j}\gamma_{2j+1}) = -\sum_j (\sigma_j^x + \sigma_j^z \sigma_{j+1}^z).
\end{equation}

\begin{figure}[htbp]
    \centering
    \includegraphics[width=0.8\linewidth]{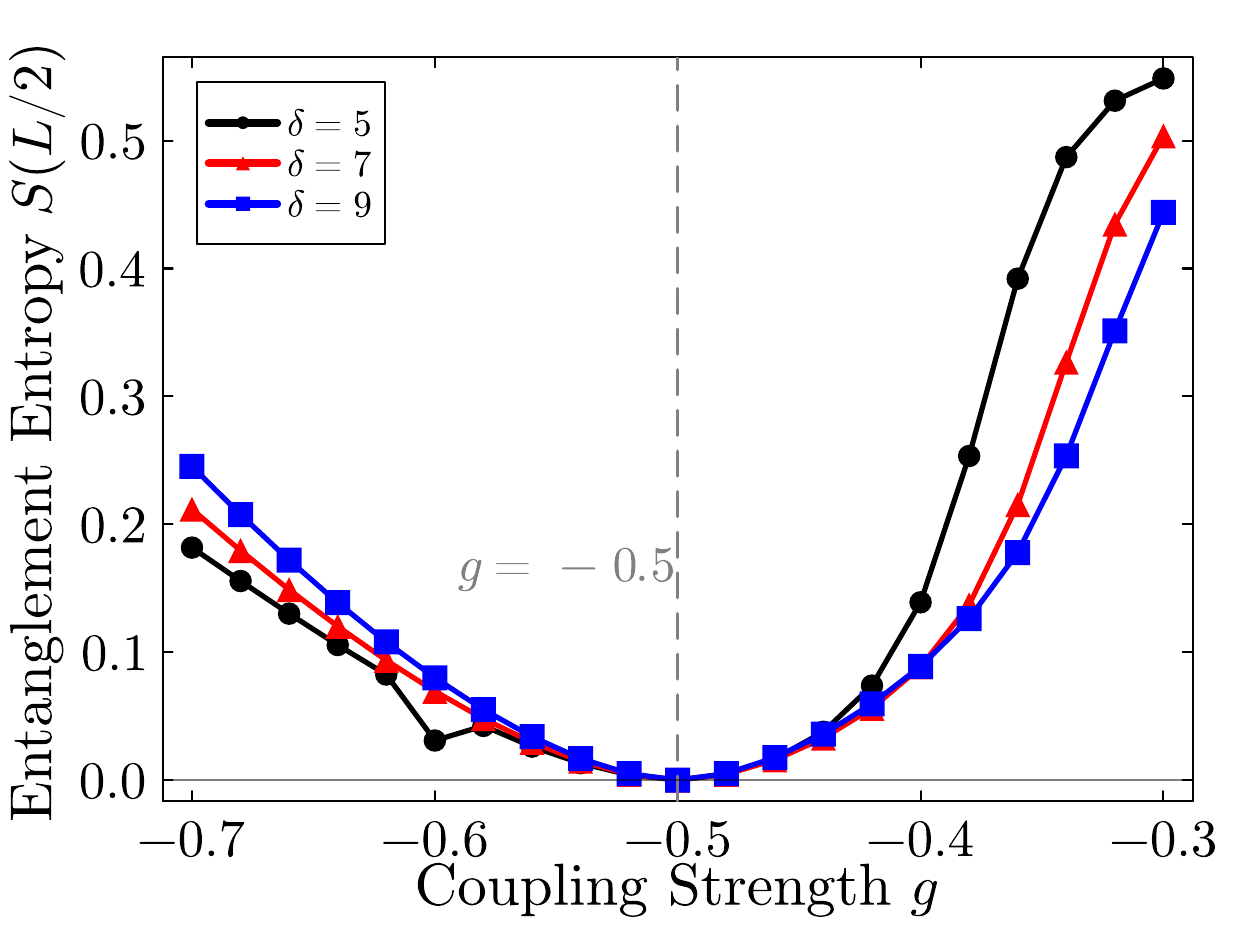} 
    \fbox{\begin{minipage}{0.8\linewidth}
    \end{minipage}}
    \caption{\label{fig:EE_dip} Numerically calculated entanglement entropy $S$ as a function of coupling $g$ for several odd interaction ranges $\delta$. A clear feature is the drop to zero entropy at $g/t = -0.5$ for all odd $\delta$ considered, indicating a shared exact product ground state at this point.}
\end{figure}

However, the form of the four-fermion interaction in the spin basis depends fundamentally on the parity of the range $\delta$. For example, the interaction term generates two distinct classes:
\begin{subequations}
\label{eq:hamiltonian_spin}
\begin{align}
\mathcal{H}_{\text{int}}^{(\text{odd }\delta)} &= -g\sum_j \left( \sigma_j^z \sigma_{j+1}^z \sigma_{j+\frac{\delta+1}{2}}^x + \sigma_j^x\sigma_{j+\frac{\delta-1}{2}}^z\sigma^z_{j+\frac{\delta+1}{2}} \right), \label{eq:H_odd}\\
\mathcal{H}_{\text{int}}^{(\text{even }\delta)} &= -g\sum_j \left( \sigma_j^z \sigma_{j+1}^z \sigma_{j+\frac{\delta}{2}}^z \sigma_{j+\frac{\delta}{2}+1}^z + \sigma_j^x \sigma_{j+\frac{\delta}{2}}^x \right), \label{eq:H_even}
\end{align}
\end{subequations}
which illustrates that the even $\delta$ series has competing interactions whereas the odd $\delta$ has a structure consistent with that of the OF\cite{obrien1} model.
It is not surprising that these models give rise to vastly different physics.   Our choice to focus on the odd $\delta$ series is justified by a remarkable property.  Namely, similar to the $\delta=3$ case \cite{obrien1}, we find that all models with odd $\delta$ share a common, exactly solvable ground state at $g/t = -0.5$ as illustrated in Fig. (\ref{fig:EE_dip}). At this coupling, the ground state becomes a simple, unentangled product state, characterized by zero entanglement entropy. This shared feature provides a stable analytical anchor point across the entire odd-$\delta$ family.  This constitutes our first substantial result in this work.  As a consequence,  we focus exclusively on the family of models with odd $\delta = 3, 5, 7, \dots$. Our primary goal is to track the shift of the TCI critical point $g_c(\delta)$ within this more tractable family.

\paragraph{1D Numerical Results as Benchmark}

To determine the phase diagram of the model, specifically the location of the TCI point, we employ DMRG calculations using the ITensor library\cite{itensor}. We perform simulations with a maximum bond dimension of $\chi=400\sim1600$ to ensure convergence.  Such a large bond dimension which is particularly crucial as the interaction range $\delta$ increases.  For moderate interaction ranges ($\delta=5, 7$), we can precisely locate the critical points by adopting the universal energy ratio method utilized by RZFA~\cite{aff1}. At play here is that at a  conformal critical point, the finite-size energy gaps scale universally according to the scaling dimensions of the underlying CFT\cite{cardyratio}. By comparing the ratios of excitation energies in different sectors, we can distinguish the TCI phase ($c=0.7$) from the standard Ising phase ($c=0.5$).  The specific universal ratios used in our analysis are summarized in Table~\ref{tab:ratios}.

\begin{table}[htbp]
  \centering
  \caption{\textbf{Universal Energy Ratios from CFT.} 
    Universal energy gap ratios for the Ising ($c=1/2$) and Tricritical Ising (TCI, $c=7/10$) conformal field theories. The states are labeled as $B_n^\tau$, where $B \in \{P, A\}$ denotes periodic or anti-periodic boundary conditions, the superscript $\tau \in \{+, -\}$ indicates the even or odd parity sector with respect to the global flip $\prod_j \sigma^x_j$, and the subscript $n \in \{0, 1\}$ refers to the ground state and the first excited state within that sector, respectively.}
  \label{tab:ratios}
  \vspace{2mm}
  \begin{tabular}{|c|c|c|c|}
    \hline
    CFT ($c$) & $\frac{A_0^- - P_0^+}{P_1^+ - P_0^+}$ & $\frac{P_0^- - P_0^+}{P_1^+ - P_0^+}$ & $\frac{P_1^- - P_0^+}{P_1^+ - P_0^+}$ \\
    \hline
    Ising (0.5) & 0.5   & 0.125 & 1.125 \\
    \hline
    TCI (0.7)   & 3.5   & 0.375 & 4.375 \\
    \hline
  \end{tabular}
\end{table}

Using this method, we successfully obtained precise phase diagrams for $\delta=5, 7$ (see Fig.~\ref{fig:supp_data1}).
However, as $\delta$ increases, the system inherently requires larger system sizes $L$ to illustrate the effects related to the interaction range, which simultaneously demands larger bond dimensions for convergence. This renders the precise ratio method computationally prohibitive for larger $\delta$. For the cases of $\delta=9, 11, 17$, we resorted to an estimation method based on entanglement entropy scaling. Based on Cardy's formula\cite{Car1} for the entanglement entropy in a CFT, $S \sim \frac{c}{6} \ln L$, and given that the TCI CFT has a central charge $c=7/10$, we can estimate $g_c$ as illustrated in Fig.~\ref{{fig:central_charge}}.  

\begin{table}[b]
    \centering
    \caption{\textbf{Critical coupling strengths $g_c$ for varying $\delta$.}}
    \label{tab:gc_delta_grid}
    \begin{tabular}{lccc}
        \toprule[1.5pt]
        $\delta(W)$ & $3(2.5)$ & $5(3.5)$ & $7(4.5)$ \\
        \midrule
        $g_c$ & $-0.428$ & $-0.370$ & $-0.335$ \\
        \midrule[1pt]
        $\delta(W)$ & $9(5.5)$ & $11(6.5)$ & $17(9.5)$ \\
        \midrule
        $g_c$ & $-0.30\sim-0.29$ & $-0.28\sim-0.27$ & $-0.25\sim-0.24$ \\
        \bottomrule[1.5pt]
    \end{tabular}
\end{table}
Our 1D numerical results reveal two key features:
\begin{enumerate}
    \item The "Ising $\to$ TCI $\to$ Gapped" phase diagram structure observed by OF\cite{obrien1} for $\delta=3$ is robust against increasing the interaction range $\delta$.
        \item The critical $|g_c|$ monotonically decreases towards zero as $\delta$ increases, as shown in Table~\ref{tab:gc_delta_grid}.
\end{enumerate}
\begin{figure}[ht]
    \centering
    \begin{minipage}[c]{1.0\linewidth}
        \centering
\stackinset{l}{3pt}{t}{3pt}{\textbf{(a)}}{%
            \includegraphics[width=\linewidth]{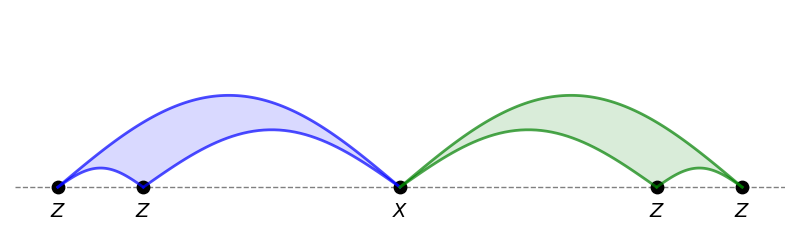}%
        }
    \end{minipage}

    \par\vspace{1em}

    \begin{minipage}[c]{0.48\linewidth}
        \centering
\stackinset{l}{3pt}{t}{3pt}{\textbf{(b)}}{%
            \includegraphics[width=\linewidth]{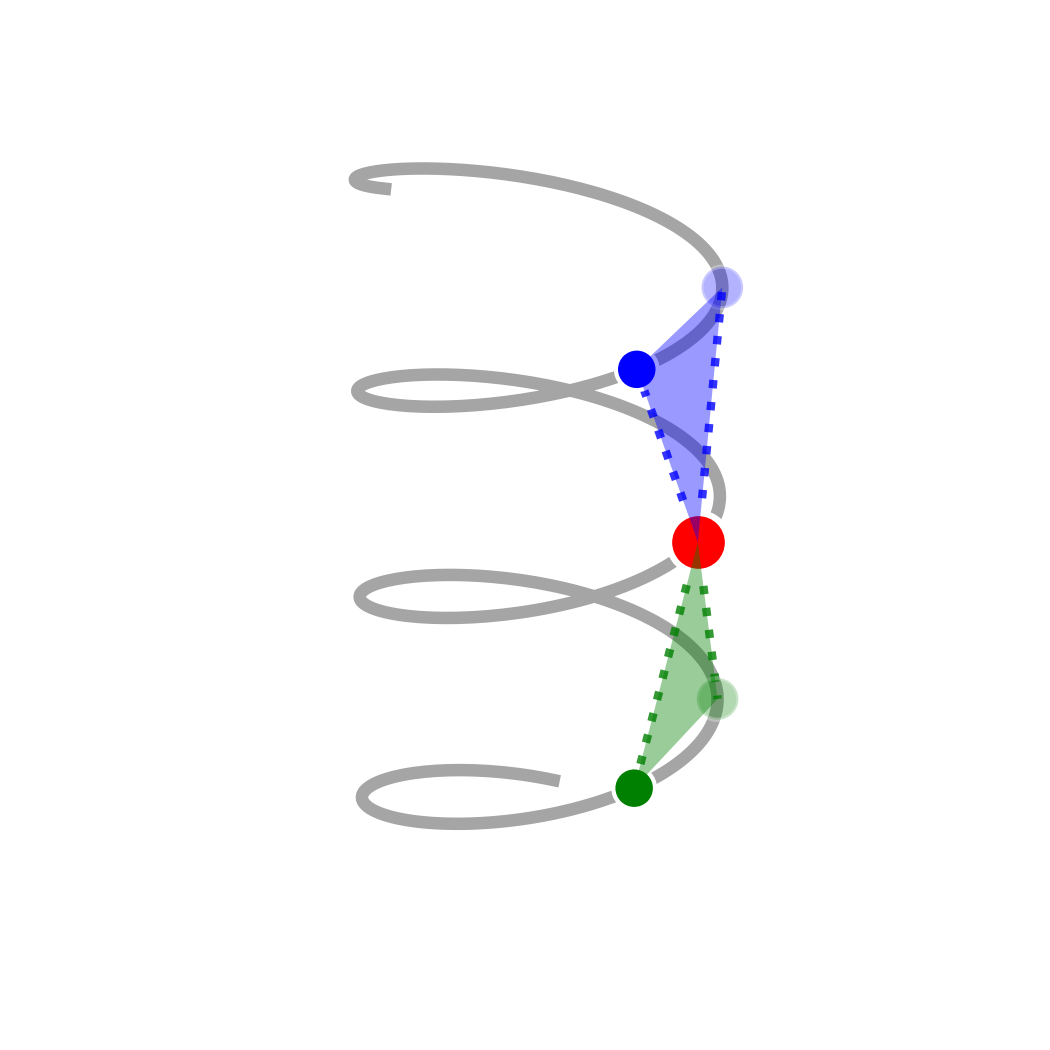}%
        }
    \end{minipage}
    \hfill
    \begin{minipage}[c]{0.4\linewidth}
        \centering
\stackinset{l}{3pt}{t}{3pt}{\textbf{(c)}}{%
            \includegraphics[width=\linewidth]{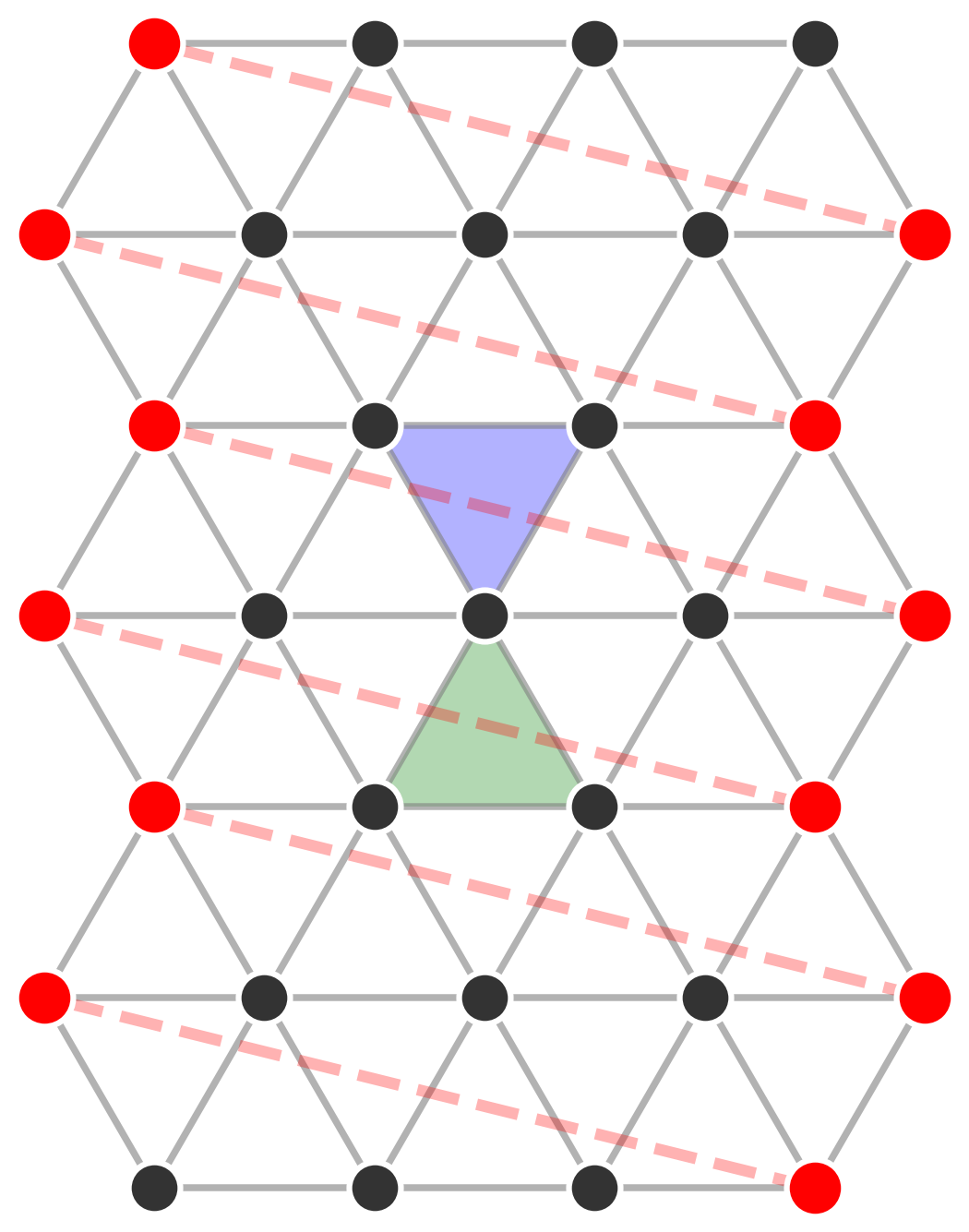}%
        }
    \end{minipage}
    \caption{\label{fig:hbc}Geometric mapping from a 1D variable-range chain to a 2D locally alike system.
(a) A 1D chain with interaction range $\delta$ as in our Hamiltonian.
(b) The corresponding 2D swirling lattice constructed by wrapping the chain helically with a period of $\frac{\delta-1}{2}$, $\frac{\delta+1}{2}\cdots$. (c) Corresponding 2D model when apply anti-helical-boundary condition to the swirling chain in (b).}
    \label{fig:three_figs}
\end{figure}
These observations naturally lead to profound questions: What is the exact functional dependence of $g_c$ on $\delta$? In the infinite-range limit ($\delta \to \infty$), do the Ising and TCI phases remain stable, or do they eventually collapse? The numerical difficulties encountered at large $\delta$ necessitate a new theoretical approach to answer these questions, which we introduce in the next section.

\paragraph{2D Finite-Size Effect from Helical Twist}

Establishing the phase diagram in the infinite-range limit ($\delta \to \infty$) presents a significant challenge for numerical methods such as DMRG, where computational costs scale prohibitively with the interaction range. To rigorously access this thermodynamic limit, we move beyond simple numerical extrapolation and establish a theoretical framework that maps our family of variable-range 1D models onto a single representative 2D local model of varying widths. This mapping enables us to interpret the $\delta$-dependent critical behavior not as an arbitrary finite-range effect, but through the well-established lens of 2D finite-size scaling (FSS)\cite{fss}.

The geometrical transformation is achieved by wrapping the 1D chain helically, a technique that fundamentally reinterprets the 1D interaction range $\delta$ as the circumference $W$ of a 2D cylinder as in Fig~\ref{fig:hbc}.
Note we recover one extra site on each row after imposing anti-helical boundary conditions.
By setting the width $W=\frac{\delta+2}{2}=\Delta/2$ (the average number of site for even and odd rows), we find that the variable-range 1D term is mapped to a strictly local nearest-neighbor bond (or gate) in the transverse ($\hat{y}$) direction of the 2D lattice.

Under this mapping, we find that the original 1D Hamiltonian is exactly equivalent to a 2D model defined on an infinite height plane of circumference $W$, subject to helical boundary conditions (HBC). The HBC, which is initially used in 2D-DMRG calculations\cite{dmrg1}, introduces a specific twist, connecting site $(W, y)$ to $(0, y+1)$, which ensures that the 1D chain topology is preserved. Crucially, as $W \to \infty$, this twist becomes negligible, and the system approaches true 2D bulk behavior.  This is the principal innovation in this work.

A crucial insight arises from this construction: in the limit of infinite chain length ($L = H \times W \to \infty$). Namely, the HBC maps the finite 2D lattice onto a single, translationally invariant 1D system. This offers a unique ''advantage'' over 2D open boundary conditions (OBC).  Namely, while the $g=0$ parallel critical TFIM chains appear gapped in 2D with OBC due to finite-size correction, the HBC provides a genuine 1D  thermodynamic limit  for realizing the gapless phase for correlations along the chain ($x$-direction in 2D).

However, the HBC cannot escape the overall 2D finite-size effect. Once the inter-chain interaction is turned on by our g-terms, the system is still subject to an energy scale imposed by the effective width (or interaction range $\delta$). Similar to OBC and PBC, the HBC cannot avoid the frustration arising from the overlap of ``correlation disks'' whose terminus lies outside the circumference of the cylinder. That is, regions of the sample that exceed the correlation length are effectively truncated.  This structural frustration, where the system energetically disfavors certain ground states or excitations, is the root of the phase transition after projection onto our 1D infinitely long system.  
Consequently, we can claim that, while we measure a true thermodynamic phase diagram for the effective 1D system, the behavior of the critical point $g_c(\delta)$ has the same root cause as that of the 2D finite-size transitions under OBC or PBC. The trajectory of $g_c$ is thus governed by the finite-size scaling (FSS) laws of the underlying (2+1)D bulk universality class.

To determine the resultant scaling laws, we first analyze the intrinsic physics of the (2+1)D bulk Hamiltonian. The model retains a Kramers-Wannier self-duality ($X \leftrightarrow ZZ$). We also observe two distinct limiting behaviors: 
\begin{enumerate}
    \item At $g=-0.5$, the system resides in a trivial, gapped paramagnetic phase with local, gapped excitations.
    \item At $g=0$, the model decouples into an infinite set of parallel, non-interacting 1D Transverse-Field Ising Models (TFIMs), each at its own critical point in the Ising Universality Class.
\end{enumerate}

\begin{figure}[htbp]
    \centering
    \includegraphics[width=\linewidth]{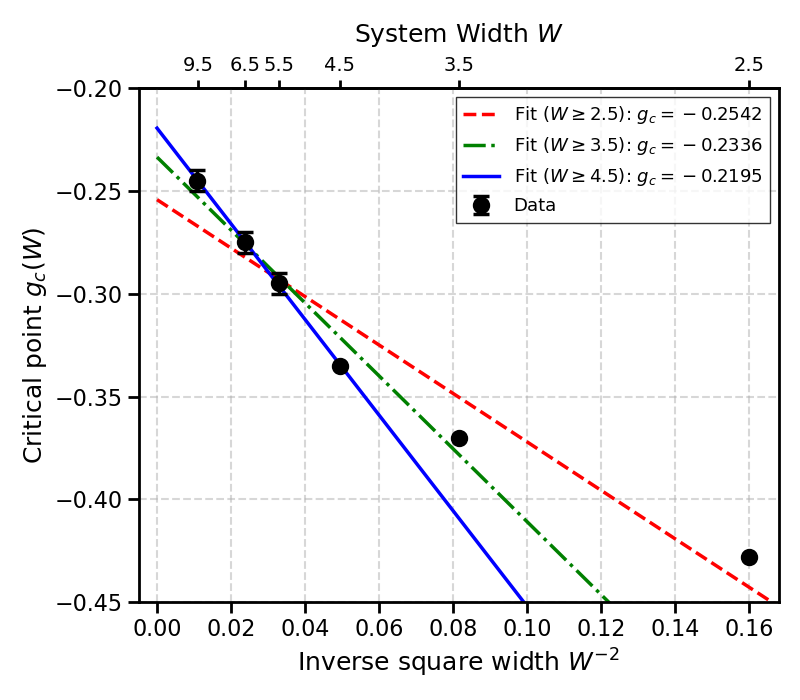}
    \caption{Linear fit of the finite size critical coupling $\mathrm{g_c}(W)$ against the scaling width $W^{-2}$ and determination of the infinitely long range limit $\mathrm{g_c}(\infty)$}
    \label{fig:benchmark_fit}
\end{figure}

The existence of a gapped phase ($g < g_c$) and a continuous critical line ($g_c<g<0$) necessitates the presence of another quantum tricritical point (QTCP) separating them.
This (2+1)D QTCP is equivalent to a 3D classical tricritical point. Such a system is described by a $\phi^6$ Landau-Ginzburg-Wilson (LGW) theory\cite{shenkar1}. The upper critical dimension $d_c$ for a $\phi^6$ theory is precisely $d_c=3$.\cite{qtcp1} 
Our (2+1)D system is therefore exactly at its upper critical dimension. In this special case ($d=d_c$), mean-field theory exponents become exact, but they are modified by multiplicative logarithmic corrections\cite{qtcp2} \cite{qtcp3}. However, as this correction emerges in higher order loop RG calculations, we can simply use power law dependence here\cite{qtcp4} of the form,
\beq
\label{eq:fss_tricritical}
g_c(\delta) = g_c(\infty) + C \cdot W^{-2} + \cdots.
\eeq
Here $g_c(\infty)$ is the bulk (2+1)D tri-critical point and $C$ is a non-universal amplitude.

Building upon this theoretical claim, we now proceed to benchmark our numerical data against the scaling law derived in Eq.~\ref{eq:fss_tricritical}. Our $g_c(\delta)$ data points, presented in Table~\ref{tab:gc_delta_grid} (entanglement or universal ratio methods), are fit to this form. This allows us to perform a robust two-parameter fit to extract the non-universal amplitude $C$ and the infinite variable-range tri-critical point $g_c \equiv g_c(\delta \to \infty)$.  From Fig.~\ref{fig:benchmark_fit}, we see that 
$g_c(\infty)$ is non-zero and roughly $-0.2$.  That this limiting value of the critical coupling is non-zero is one of our key results. 

This data is inevitably contaminated by non-universal contributions from higher-order, irrelevant scaling terms that are not captured in Eq.~\ref{eq:fss_tricritical}. Nevertheless, as illustrated in Fig.~\ref{fig:benchmark_fit}, the theoretical curve provides a collapse to the numerical data points. The agreement between the predicted scaling law and the benchmark data serves as a strong consistency check. 

\paragraph{Conclusion and Outlook}

In this work, we have systematically investigated a family of finite-range interacting Majorana fermion models. By focusing on the odd interaction range series ($\delta=3, 5, \dots$), we identified a robust class of lattice realizations of TCI conformal field theories. These models share the same universal critical physics.

Our numerical results reveal a monotonic drift of the TCI critical point $|g_c(\delta)|$ towards smaller values as the interaction range increases. To overcome the severe computational challenges in the large-$\delta$ regime, we introduced a novel theoretical framework that maps these variable range 1D chains onto locally similar 2D lattice models with helical boundary conditions. This mapping proved to be more than just a computational convenience; rather, it provides a powerful physical bridge. By applying 2D finite-size scaling theory to our 1D data, we discovered a connection between lower-dimensional finite range and its hidden higher dimensional geometrical structure.

Looking forward, this 1D-2D embedding framework holds significant promise for broader applications. An immediate next step is to apply this perspective to the more challenging even-$\delta$ series. It is plausible that the "frustration" observed in 1D for even $\delta$ maps onto geometric frustration in the corresponding 2D lattice, offering a new geometric intuition for their complexity. More generally, this methodology could be adapted to other classes of 1D chains with long-range interactions, potentially unveiling hidden higher-dimensional structures in a wide range of ostensibly quantum many-body systems.
\newline

\vfill

\bibliographystyle{apsrev4-2}
\bibliography{my_ref}

@article{tci1,
  title = {Ising Model for the $\ensuremath{\lambda}$ Transition and Phase Separation in ${\mathrm{He}}^{3}$-${\mathrm{He}}^{4}$ Mixtures},
  author = {Blume, M. and Emery, V. J. and Griffiths, Robert B.},
  journal = {Phys. Rev. A},
  volume = {4},
  issue = {3},
  pages = {1071--1077},
  numpages = {0},
  year = {1971},
  month = {Sep},
  publisher = {American Physical Society},
  doi = {10.1103/PhysRevA.4.1071},
  url = {https://link.aps.org/doi/10.1103/PhysRevA.4.1071}
}

@article{shenkar1,
title = {RG flow in N = 1 discrete series},
journal = {Nuclear Physics B},
volume = {316},
number = {3},
pages = {590-608},
year = {1989},
issn = {0550-3213},
doi = {https://doi.org/10.1016/0550-3213(89)90060-6},
url = {https://www.sciencedirect.com/science/article/pii/0550321389900606},
author = {D.A. Kastor and E.J. Martinec and S.H. Shenker}
}

@article{aff1,
  title = {Emergent Supersymmetry from Strongly Interacting Majorana Zero Modes},
  author = {Rahmani, Armin and Zhu, Xiaoyu and Franz, Marcel and Affleck, Ian},
  journal = {Phys. Rev. Lett.},
  volume = {115},
  issue = {16},
  pages = {166401},
  numpages = {5},
  year = {2015},
  month = {Oct},
  publisher = {American Physical Society},
  doi = {10.1103/PhysRevLett.115.166401},
  url = {https://link.aps.org/doi/10.1103/PhysRevLett.115.166401}
}

@article{obrien1,
  title = {Lattice Supersymmetry and Order-Disorder Coexistence in the Tricritical Ising Model},
  author = {O'Brien, Edward and Fendley, Paul},
  journal = {Phys. Rev. Lett.},
  volume = {120},
  issue = {20},
  pages = {206403},
  numpages = {5},
  year = {2018},
  month = {May},
  publisher = {American Physical Society},
  doi = {10.1103/PhysRevLett.120.206403},
  url = {https://link.aps.org/doi/10.1103/PhysRevLett.120.206403}
}

@article{itensor,
	title={{The ITensor Software Library for Tensor Network Calculations}},
	author={Matthew Fishman and Steven R. White and E. Miles Stoudenmire},
	journal={SciPost Phys. Codebases},
	pages={4},
	year={2022},
	publisher={SciPost},
	doi={10.21468/SciPostPhysCodeb.4},
	url={https://scipost.org/10.21468/SciPostPhysCodeb.4}
}

@article{Car1,
   title={Entanglement entropy and quantum field theory},
   volume={2004},
   ISSN={1742-5468},
   url={http://dx.doi.org/10.1088/1742-5468/2004/06/P06002},
   DOI={10.1088/1742-5468/2004/06/p06002},
   number={06},
   journal={Journal of Statistical Mechanics: Theory and Experiment},
   publisher={IOP Publishing},
   author={Pasquale Calabrese and John Cardy},
   year={2004},
   month=jun, pages={P06002} }

@article{qtcp1,
  title = {Scaling Approach to Tricritical Phase Transitions},
  author = {Riedel, Eberhard K.},
  journal = {Phys. Rev. Lett.},
  volume = {28},
  issue = {11},
  pages = {675--678},
  numpages = {0},
  year = {1972},
  month = {Mar},
  publisher = {American Physical Society},
  doi = {10.1103/PhysRevLett.28.675},
  url = {https://link.aps.org/doi/10.1103/PhysRevLett.28.675}
}

@article{qtcp2,
  title = {Tricritical Exponents and Scaling Fields},
  author = {Riedel, Eberhard K. and Wegner, Franz J.},
  journal = {Phys. Rev. Lett.},
  volume = {29},
  issue = {6},
  pages = {349--352},
  numpages = {0},
  year = {1972},
  month = {Aug},
  publisher = {American Physical Society},
  doi = {10.1103/PhysRevLett.29.349},
  url = {https://link.aps.org/doi/10.1103/PhysRevLett.29.349}
}

@article{qtcp3,
  title = {Logarithmic Corrections to the Molecular-Field Behavior of Critical and Tricritical Systems},
  author = {Wegner, Franz J. and Riedel, Eberhard K.},
  journal = {Phys. Rev. B},
  volume = {7},
  issue = {1},
  pages = {248--256},
  numpages = {0},
  year = {1973},
  month = {Jan},
  publisher = {American Physical Society},
  doi = {10.1103/PhysRevB.7.248},
  url = {https://link.aps.org/doi/10.1103/PhysRevB.7.248}
}

@article{qtcp4,
  title = {Logarithmic corrections to the mean-field theory of tricritical points},
  author = {Stephen, Michael J. and Abrahams, Elihu and Straley, Joseph P.},
  journal = {Phys. Rev. B},
  volume = {12},
  issue = {1},
  pages = {256--262},
  numpages = {0},
  year = {1975},
  month = {Jul},
  publisher = {American Physical Society},
  doi = {10.1103/PhysRevB.12.256},
  url = {https://link.aps.org/doi/10.1103/PhysRevB.12.256}
}

@article{dmrg1,
  title = {One-dimensional projection of two-dimensional systems using spiral boundary conditions},
  author = {Kadosawa, Masahiro and Nakamura, Masaaki and Ohta, Yukinori and Nishimoto, Satoshi},
  journal = {Phys. Rev. B},
  volume = {107},
  issue = {8},
  pages = {L081104},
  numpages = {5},
  year = {2023},
  month = {Feb},
  publisher = {American Physical Society},
  doi = {10.1103/PhysRevB.107.L081104},
  url = {https://link.aps.org/doi/10.1103/PhysRevB.107.L081104}
}

@article{cardyratio,
  title = {Conformal invariance, the central charge, and universal finite-size amplitudes at criticality},
  author = {Bl\"ote, H. W. J. and Cardy, John L. and Nightingale, M. P.},
  journal = {Phys. Rev. Lett.},
  volume = {56},
  issue = {7},
  pages = {742--745},
  numpages = {0},
  year = {1986},
  month = {Feb},
  publisher = {American Physical Society},
  doi = {10.1103/PhysRevLett.56.742},
  url = {https://link.aps.org/doi/10.1103/PhysRevLett.56.742}
}

@article{fss,
title = {Finite size effects in phase transitions},
journal = {Nuclear Physics B},
volume = {257},
pages = {867-893},
year = {1985},
issn = {0550-3213},
doi = {https://doi.org/10.1016/0550-3213(85)90379-7},
url = {https://www.sciencedirect.com/science/article/pii/0550321385903797},
author = {E. Brézin and J. Zinn-Justin}
}

\pagebreak
\widetext
\begin{center}
\textbf{\large Supplemental Material: 2D Helical Twist Controls  Tricritical Point in an Interacting Majorana Chain}
\end{center}
\setcounter{equation}{0}
\setcounter{figure}{0}
\setcounter{table}{0}
\setcounter{page}{1}
\makeatletter
\renewcommand{\theequation}{S\arabic{equation}}
\renewcommand{\thefigure}{S\arabic{figure}}
\renewcommand{\bibnumfmt}[1]{[S#1]}
\renewcommand{\citenumfont}[1]{S#1}
\makeatother

In this Supplemental Material, we present detailed numerical evidence supporting the robustness of the phase diagram for larger interaction ranges and get accurate or approximate value of $g_c$.

\section{S-I. Universal Spectrum Ratios}

\begin{figure}[h]
    \centering
    \begin{minipage}{0.48\textwidth}
        \centering
        \includegraphics[width=\linewidth, height=5cm]{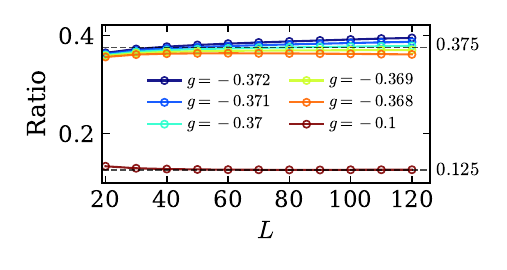}
    \end{minipage}
    \hfill
    \begin{minipage}{0.48\textwidth}
        \centering
        \includegraphics[width=\linewidth, height=5cm]{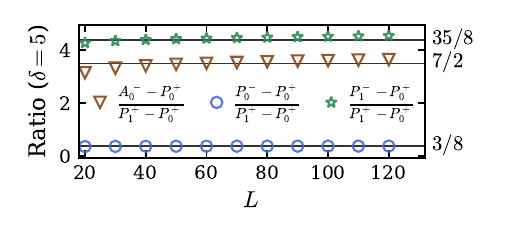}
    \end{minipage}

    \vspace{0.5cm}

    \begin{minipage}{0.48\textwidth}
        \centering
        \includegraphics[width=\linewidth, height=5cm]{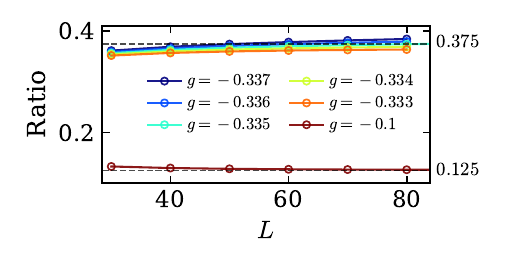}
    \end{minipage}
    \hfill
    \begin{minipage}{0.48\textwidth}
        \centering
        \includegraphics[width=\linewidth, height=5cm]{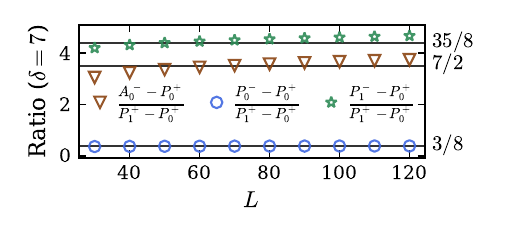}
    \end{minipage}
    
\caption{\label{fig:supp_data1}
Finite-size scaling analysis of the spectrum ratios for $\delta=5$ (top row) and $\delta=7$ (bottom row). The left panels show the scan of the selected ratio versus coupling $g$, while the right panels show the comparison of three different ratios against TCI CFT predictions at the determined critical points.
}
\end{figure}

In the left column of Fig.~\ref{fig:supp_data1}, we perform a fine-grained scan (interval $0.001$) of the ratio $\frac{P_0^- - P_0^+}{P_1^+ - P_0^+}$.
The data clearly illustrate that for small $|g|$, the scaling of the system tends towards the Ising CFT limit, while near the critical point $g_c$, the ratio exhibits a remarkably flat dependence on the size of the system $L$, closely matching the prediction of tricritical Ising (TCI).
In the right column, we compare all three ratios (as defined in Table~I of the main text) with the theoretical CFT values at the identified critical points; all three show excellent agreement.
These results confirm that the ``Ising CFT ($|g| < |g_c|$) $\to$ TCI CFT ($g \approx g_c$)'' behavior persists for these $\delta$ values.
Combined with the analysis of the trivial gapped state at $g=-0.5$ in the main text, this establishes the robustness of the phase diagram. And we can also yield the critical points $g_c(\delta=5) = -0.370$ and $g_c(\delta=7) = -0.335$.
\clearpage

\section{S-II. Subtracted Entanglement Entropy}

To extend our analysis to larger interaction ranges ($\delta=9, 11, 17$), we employ the Cardy-Calabrese formula discussed in the main text to approximate the location of the critical point $g_c$.
This approach provides an estimation with a precision of approximately $0.1$, establishing a crucial numerical benchmark for subsequent theoretical predictions.
Notably, by comparing these estimated values with the precise results from the earlier section, we observe a trend where the magnitude of the critical point $|g_c|$ decreases towards $0$ as the interaction range $\delta$ increases.

\begin{figure}[h]
    \centering
    \begin{minipage}{0.65\textwidth}
        \centering
        \includegraphics[width=\linewidth, height=5cm]{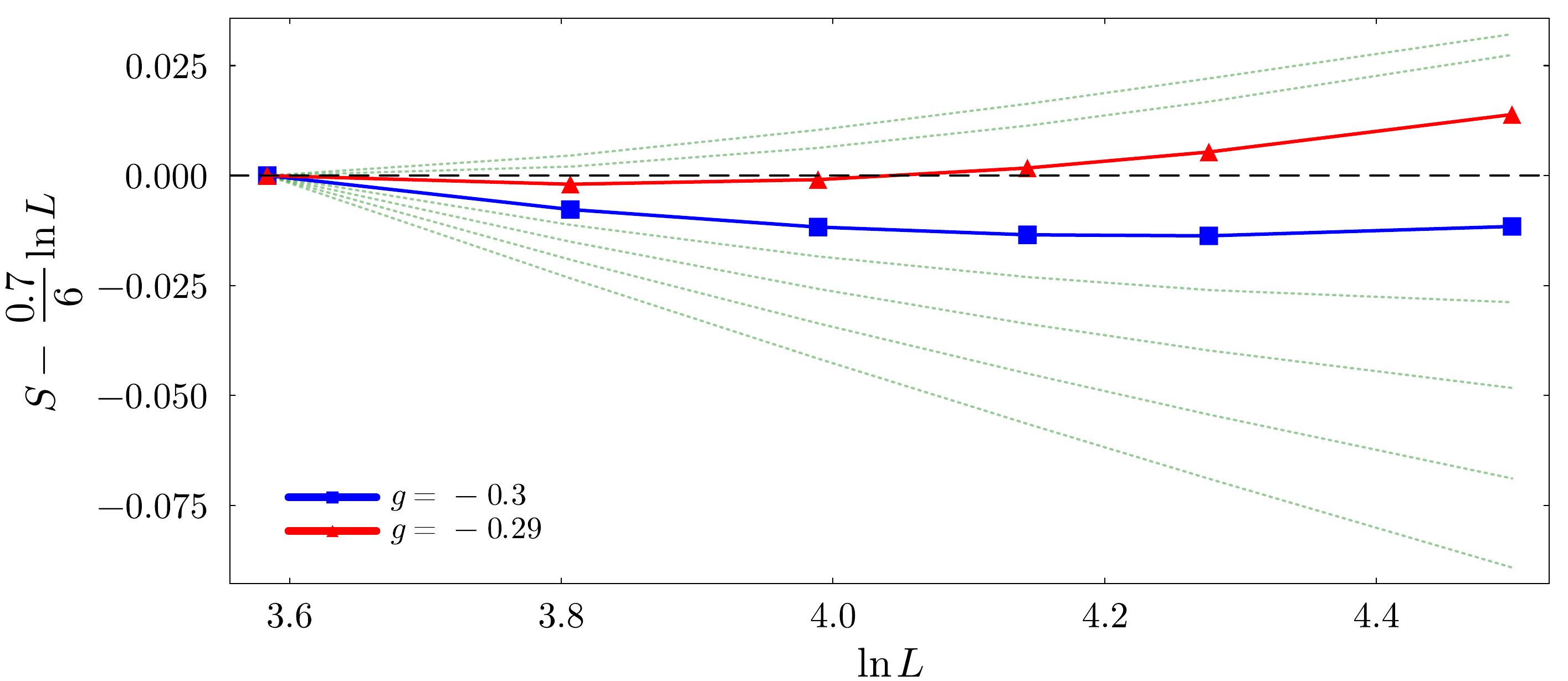}
    \end{minipage}\\[0.6cm]
    
    \begin{minipage}{0.65\textwidth}
        \centering
        \includegraphics[width=\linewidth, height=5cm]{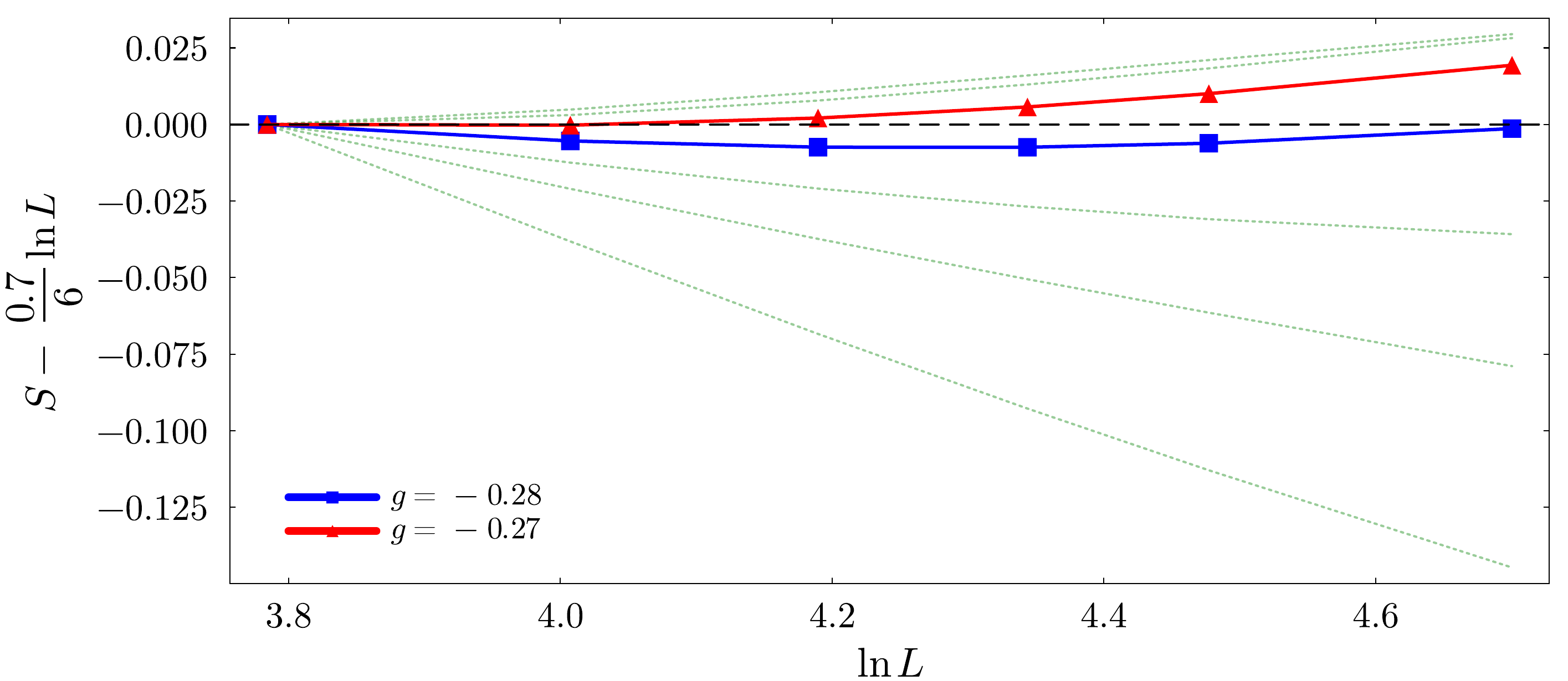}
    \end{minipage}\\[0.6cm]
    
    \begin{minipage}{0.65\textwidth}
        \centering
        \includegraphics[width=\linewidth, height=5cm]{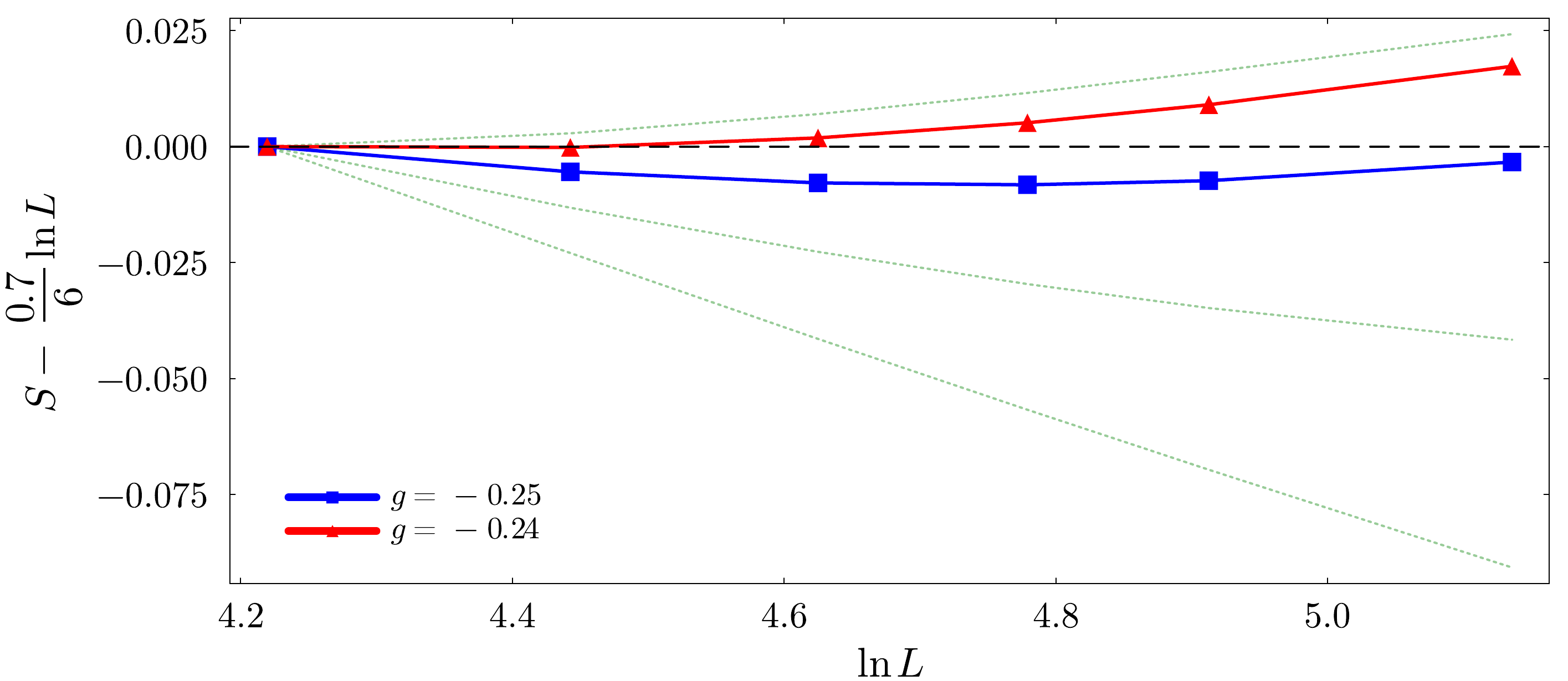}
    \end{minipage}
    
\caption{\label{{fig:central_charge}}
Diagnostic plot for the central charge approximation. 
By plotting the subtracted entanglement entropy $S(L) - \frac{0.7}{6}\ln L$, we highlight the estimated bounds of the critical point, represented by the red (upper) and blue (lower) curves. 
The flat dashed line marks the theoretical reference value ($c=0.7$).
The background green lines correspond to other interaction strengths $g$ with a separation of 0.1.
}
\end{figure}

\end{document}